# ML-Enabled Systems Model Deployment and Monitoring: Status Quo and Problems

Eduardo Zimelewicz[1], Marcos Kalinowski[1], Daniel Mendez[2,9], Görkem Giray[3], Antonio Pedro Santos Alves[1], Niklas Lavesson[2], Kelly Azevedo[1], Hugo Villamizar[1], Tatiana Escovedo[1], Helio Lopes[1], Stefan Biffl[4], Juergen Musil[4], Michael Felderer[5,6], Stefan Wagner[7], Teresa Baldassarre[8], and Tony Gorschek[2,9]

[1] Pontifical Catholic University of Rio de Janeiro (PUC-Rio), Brazil
[2] Blekinge Institute of Technology (BTH), Sweden
[3] Independent Researcher, Turkey
[4] Vienna University of Technology (TU Wien), Austria
[5] German Aerospace Center (DLR), Germany
[6] University of Cologne, Germany
[7] University of Stuttgart, Germany
[8] University of Bari, Italy
[9] fortiss GmbH, Germany

**Abstract.** [Context] Systems that incorporate Machine Learning (ML) models, often referred to as ML-enabled systems, have become commonplace. However, empirical evidence on how ML-enabled systems are engineered in practice is still limited; this is especially true for activities surrounding ML model dissemination. [Goal] We investigate contemporary industrial practices and problems related to ML model dissemination, focusing on the model deployment and the monitoring ML life cycle phases. [Method] We conducted an international survey to gather practitioner insights on how ML-enabled systems are engineered. We gathered a total of 188 complete responses from 25 countries. We analyze the status quo and problems reported for the model deployment and monitoring phases. We analyzed contemporary practices using bootstrapping with confidence intervals and conducted qualitative analyses on the reported problems applying open and axial coding procedures. [Results] Practitioners perceive the model deployment and monitoring phases as relevant and difficult. With respect to model deployment, models are typically deployed as separate services, with limited adoption of MLOps principles. Reported problems include difficulties in designing the architecture of the infrastructure for production deployment and legacy application integration. Concerning model monitoring, many models in production are not monitored. The main monitored aspects are inputs, outputs, and decisions. Reported problems involve the absence of monitoring practices, the need to create custom monitoring tools, and the selection of suitable metrics. [Conclusion] Our results help provide a better understanding of the adopted practices and problems in practice and support guiding ML deployment and monitoring research in a problem-driven manner.

# 1 Introduction

In recent years, the advancements in Machine Learning (ML) and, altogether, Artificial Intelligence (AI), have helped the incoming of technological innovation and transformation across various industries. These ML-enabled systems have shown capabilities in automating complex tasks, making data-driven decisions, and enhancing overall efficiency. However, despite their immense potential, the implementation of ML-enabled systems requires practitioners to adapt processes to successfully develop, deploy, and monitor in production operation. In the same level, Software Engineering (SE) practices can help to speed up the development of such features. However, ML-enabled systems are inherently different by nature affecting rendering traditional SE practices insufficient to be directly applied, thus, revealing new challenges [1].

In regard to the current increase in ML system usage, this paper aims to identify potential industrial problems and the current status quo in terms of practices applied in the development of ML-enabled software systems. With the main goal of understanding the pain points and how those systems are made, we conducted a questionnaire-based online survey. Although many other concerns appeared in the responses, such as issues in Requirements Engineering and Data Quality, the work presented in this paper focuses on the model deployment and monitoring of ML-enabled systems. Our focus is on evaluating exprienced challenges as well as approaches employed.

The main findings show that practitioners perceive the model deployment and monitoring phases as relevant, but also challenging. With respect to model deployment, we observed that models are mainly deployed as separate services and that embedding the model within the consuming application or platform-as-a-service solutions are less frequently explored. Most practitioners do not follow MLOps principles and do not have an automated pipeline to retrain and rede- ploy the models, where the reported deployment problems include difficultiesin designing the architecture of the infrastructure for production, considering scalability and financial constraints, and legacy application integration. Concerning model monitoring, many of the models in production are not monitored at all, with the main aspects in scope of monitored are being outputs and decisions taken. Reported problems include not having model appropriate monitoring practices in place, the need for developing customized monitoring tools, and difficulties choosing the appropriate metrics.

The remainder of this paper is organized as follows. Section 2 provides the background and related work. In Section 3, we describe the research method. Section 4 presents then the results which we discuss further in Section 5. In section 6, we critically reflect upon the threats to validity and mitigation actions before concluding our paper with Section 7.

## 2 Background and Related Work

Machine Learning (ML) has witnessed various advancements in recent years, transforming various industries by enabling intelligent decision-making systems. Deploying ML models into real-world applications, however, presents complex challenges related to model performance, reliability, and maintenance. This section provides an overview of the research landscape concerning the deployment and monitoring of Machine Learning systems.

The use of Machine Learning in practical applications dates back to the year of 1952 when English Mathematician Arthur Samuel created the first Machine Learning program to play championship-level game of checkers [2]. However, it is in the past decade that ML deployments have gained widespread attention in practice due to the availability of large datasets, more powerful computing hardware, and improved algorithms. Despite the rapid growth in ML adoption, there still exists a significant gap between the development of ML models in testing environments and their successful deployment in real-world settings, as reported by Paleyes *et. al.* [3], especially in the fields of integration, monitoring, and updating a model. Further discussions show that, within the model deployment phase (which includes the monitoring part), adapting existing techniques such as DevOps could be extremely helpful to make development and production environments even closer, where the term MLOps follows the same concept by bringing together data scientists and operations teams, with Meenu *et. al.* [4] identifying the activities and placing stages, by conducting a systematic literature review (SLR) and grey literature review (GLR), in which organizations can improve their MLOps adoption.

To represent the main issues to transition models to production architectures, some challenges were also identified and categorized by Lewis *et. al.* [5] in four spaces. First, utilizing software architecture practices that are proven effective to traditional applications, but do not take into account the data driven aspect of such projects, meaning that the design and development of ML models, will have to be approached with new frameworks, as the one presented by Meenu *et. al.* [6] Second, creating patterns and tactics to achieve ML Quality Attributes (QAs), where existing metrics will need to be revisited and new ones will be created to better evaluate systems. Third, the monitorability as a driving quality attribute, by having the infrastructure behind the monitoring platform to be responsible for collecting specific information related to changes in the dataset, as well as the incorporated user feedback, to observe the impacts to deployed ML systems. Fourth, co-architecting and co-versioning, where the architecture of the ML system itself, alongside the architecture that supports its life cycle, will have to be developed in sync, like the MLOps pipeline and the system integration, and the existing dataset as well as the programming code.

Apart from the architecture challenges, previous research has explored different deployment models for ML systems, as the SLR, and a GLR, conducted by Meenu *et. al.* [7] by providing an overview of the AI deployment's status quo and practices to further design a deployment framework for these systems. Today's approaches range from traditional batch processing [8] to real-time streaming deployments [9] and, most currently, an increase in use of the cloud service offerings such as FaaS (Function as a Service) [10], SaaS (Software as a Service) [11], PaaS (Platform as a Service) [12] and IaaS (Infrastructure as a Service) [13], representing the benefits of cloud adoption by the practitioners such as the relief from the burden of servers' management, faster time to go into production, cost optimization and performance increase. Alongside the deployment models, the existing software architectures approaches are also getting adapted to ML models such as containerization [14], microservices [15], and serverless computing [16] have gained prominence in ensuring model deployment flexibility and scalability.

Recent studies have focused on the monitoring and maintenance of ML models. Researchers have proposed techniques for detecting Machine Learning specific metrics such as model drift, handling concept drift, and ensuring that models remain accurate and reliable over time [17, 18], which involves concepts such as statistical process control, anomaly detection, and continuous integration/continuous deployment (CI/CD) practices.

The presented literature demonstrates the diverse nature of ML deployment and monitoring challenges. While numerous strategies and techniques have been proposed, there remains a need for a holistic framework that addresses these challenges and their current approach to solve them. Building upon the insights gained from the review of existing literature and the applied survey, this paper presents an overview to address the challenges of ML model deployments and monitoring. In the subsequent sections, we delve into the details of our research survey.

## 3 Research Method

### 3.1 Goal and Research Questions

The main goal of the research study focused on surveying the current status quo and problems through the entire development lifecycle of a ML system, but for the context of the current paper, the analysis will be based on two of the most problematic concerns in maintaining the model: (i) making the model available as quickly as possible in production and (ii) managing the model and re-training it along its continuous deployment based on monitored aspects. From this goal, we inferred the following research questions:

– RQ1. What are contemporary practices for deploying ML models?
  Under this question, we aim at identifying the in-use practices and trends of the deployment stage and can refine it further into three more detailed questions:
  – RQ1.1. What kind of approaches are used to deploy ML models?
  – RQ1.2. Which tools are used for automating model retraining?
  – RQ1.3. What are the MLOps practices and principles used?

- RQ2. What are the main problems faced during the deployment in the ML life cycle stage?
- RQ3. What are contemporary practices for monitoring ML models?
  Under this question, we aim at identifying the in-use practices and trends of the *monitoring* stage and can refine itinto two more detailed questions:
  - RQ3.1. What percentage of the ML-enabled system projects that get deployed into production have their ML models actually being monitored?
  - RQ3.2. What aspects of the models are monitored?
- RQ4. What are the main problems faced during the monitoring in the ML life cycle stage?
- RQ5. What is the percentage of projects that effectively go into production?

### 3.2 Survey Design

We designed our survey based on best community practices of survey research [19], carefully conducting, in essence, the following steps:

- **Step 1. Initial Survey Design**. We conducted a literature review on ML deployment and monitoring and combined our findings with previous results on problems and the status quo to provide the theoretical foundations for questions and answer options. From there, we drafted the initial survey by involving Software Engineering and Machine Learning researchers of PUC-Rio (Brazil) with experience in R&D projects involving ML-enabled systems.
- **Step 2. Survey Design Review**. The survey was reviewed and adjusted based on online discussions and annotated feedback from Software Engineeringand Machine Learning researchers of BTH (Sweden). Thereafter, the survey was also reviewed by the other co-authors.
- **Step 3. Pilot Face Validity Evaluation**. This evaluation involves a lightweight review by randomly chosen respondents. It was conducted with 18 Ph.D. students taking a Survey Research Methods course at UCLM (Spain) taught by the second author. They were asked to provide feedback on the clearness of the questions and to record their response time. This phase re- sulted in minor adjustments related to usability aspects and unclear wording. The answers were discarded before launching the survey.
- **Step 4. Pilot Content Validity Evaluation**. This evaluation involves sub- ject experts from the target population. Therefore, we selected five experi- enced data scientists developing ML-enabled systems, asked them to answer the survey, and gathered their feedback. The participants had no difficulties answering the survey, and it took an average of 20 minutes. After this step, the survey was considered ready to be launched.

The final survey started with a consent form describing the purpose of the study and stating that it is conducted anonymously. The remainder was divided into 15 demographic questions (D1 to D15) followed by three specific parts with 17 substantive questions (Q1 to Q17): 7 on the ML life cycle and problems, 5 on requirements, and 5 on deployment and monitoring. This paper focuses on the ML life cycle problems related to model deployment and aspects of monitoring, and the specific questions regarding problems motives. The excerpts of the questions we deem relevant in context of the paper at hands are shown in Table 1. The survey was implemented using the Unipark Enterprise Feedback Suite.

## 3.3 Data Collection

Our target population concerns professionals involved in building ML-enabled systems, including different activities, such as management, design, and development. Therefore, it includes practitioners in positions such as project leaders, requirements engineers, data scientists, and developers. We used convenience sampling, sending the survey link to professionals active in our partner companies, and also distributed it openly on social media. We excluded participants that informed having no experience with ML-enabled system projects. Data collection was open from January 2022 to April 2022. In total, we received responsesfrom 276 professionals, out of which 188 completed all four survey sections. The average time to complete the survey was 20 minutes. We conservatively considered only the 188 fully completed survey responses.

## 3.4 Data Analysis Procedures

For data analysis purposes, given that all questions were optional, the number of responses varies across the survey questions. Therefore, we explicitly indicate the number of responses when analyzing each question.

Research questions *RQ1.1, RQ3.1, RQ3.2, and RQ5* concern a mix of closed questions and optional free fields, so we decided to use inferential statistics to analyze them. Our population has an unknown theoretical distribution (*i.e.*, the distribution of ML-enabled system professionals is unknown). In such cases, re-sampling methods - like bootstrapping - have been reported to be more reliable and accurate than inference statistics from samples [20, 19]. Hence, we use boot-strapping to calculate confidence intervals for our results, similar as done in [21]. In short, bootstrapping involves repeatedly taking samples with replacements and then calculating the statistics based on these samples. For each question,we take the sample of $n$ responses for that question and bootstrap $S$ resamples (with replacements) of the same size $n$. We assume $n$ as the total valid answers of each question [22], and we set 1000 for $S$, which is a value that is reported to allow meaningful statistics [23].

For research questions *RQ1.2, RQ1.3, RQ2, RQ3.1 and RQ4*, which seek to identify the main problems faced by practitioners involved in engineering ML-enabled systems related to model deployment and monitoring, alongside questions regarding which current practices are being applied, what amount of models that are generally available for users and the current monitored aspects, had their corresponding survey question designed to be open text. We conducted a qualitative analysis using open and axial coding procedures from grounded theory [24] to allow the problems to emerge from the open-text responses reflecting the experience of the practitioners. The qualitative coding procedures were conducted by one PhD student, reviewed by her advisor at PUC-Rio, and reviewed independently by three researchers from two additional sites (two from BTH Sweden and one independent researcher from Turkey). The questionnaire, the collected data, and the quantitative and qualitative data analysis artifacts, including Python scripts for the bootstrapping statistics and graphs and the peer-reviewed qualitative coding spreadsheets, are available in our open science repository [1].

**Table 1.** Research questions and survey questions

| RQ | Survey No. | Description | Type |
|---|---|---|---|
| - | ... | ... | ... |
| RQ5 | D7 | How many ML-enabled system projects have you participated in? Please, provide your best estimate: | Open |
| RQ5 | D8 | Of all the ML-enabled system projects you have participated in, how many were actually deployed into a production environment (e.g., released to the final customer)? Please, provide your best estimate: | Open |
| - | ... | ... | ... |
| RQ2 | Q4 | According to your personal experience, please outline the main problems or difficulties (up to three) faced during each of the seven ML life cycle stages. | Open |
| RQ4 | Q4 | According to your personal experience, please outline the main problems or difficulties (up to three) faced during each of the seven ML life cycle stages. | Open |
| - | ... | ... | ... |
| RQ1.1 | Q13 | In the context of the ML-enabled system projects you participatedin, which approach is typically used to deploy ML models? | Multiple Option and Free Field |
| RQ1.2 | Q14 | Do you/your organization follow the practice and principles of ML-Ops in ML-enabled system projects? For instance, do you have an automated pipeline to retrain and deploy your ML models? | Single Option and Free Field |
| RQ1.3 | Q14 | Do you/your organization follow the practice and principles of ML-Ops in ML-enabled system projects? For instance, do you have an automated pipeline to retrain and deploy your ML models? | Single Option and Free Field |
| RQ3.1 | Q15 | Based on your experience, what percentage of the ML-enabled system projects that get deployed into production have their MLmodels actually being monitored? | Open |
| RQ3.2 | Q16 | Which of the following ML model aspects are monitored for the deployed ML-enabled system projects you have worked on? | Multiple Option and Free Field |
| - | ... | ... | ... |

## 4  Results

All of the data that follows the study come with the bootstrapped samples together with the 95% confidence interval. The $N$ in each figure caption is the number of participants that answered this question. We report the proportion $P$ of the participants that checked the corresponding answer and its 95% confidence interval in square brackets.

### 4.1  Study Population

Figure 1 summarizes demographic information on the survey participants' countries, roles, and experience with ML-enabled system projects in years. It is possible to observe that the participants came from different parts of the world, representing various roles and experiences. While the figure shows only the ten countries with the most responses, we had respondents from 25 countries. As expected, our convenience sampling strategy influenced the countries, with most responses being from diverse countries (Brazil, Turkey, Austria, Germany, Sweden, and Italy).

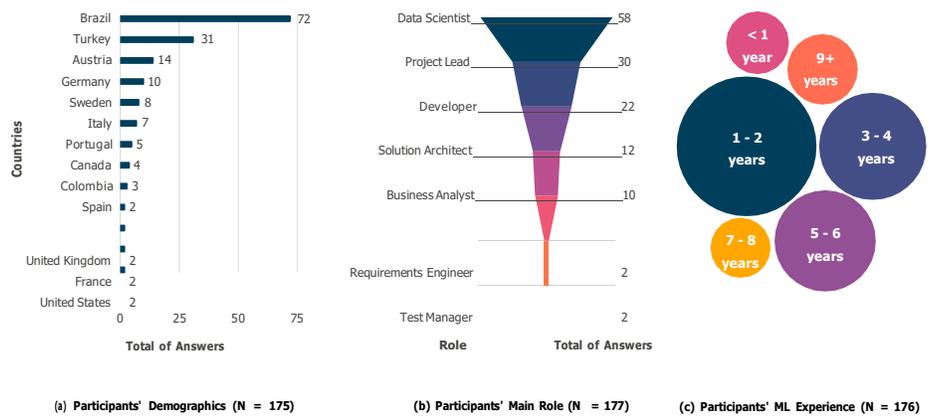

(a) Participants' Demographics (N = 175)   (b) Participants' Main Role (N = 177)   (c) Participants' ML Experience (N = 176)

**Fig. 1.** Demographic graphs for participant's countries, roles and ML work experience

Regarding employment, 45% of the participants are employed in large companies (2000+ employees), while 55% work in smaller ones of different sizes. It is possible to observe that they are mainly data scientists, followed by project leaders, developers, and solution architects. Regarding their experience with ML-enabled systems, most of the participants reported having 1 to 2 years of experience. Following closely, another substantial group of participants indicated a higher experience bracket of 3 to 6 years. This distribution highlights a balanced representation of novice and experienced practitioners. Regarding the participants' educational background, 81.38% mentioned having a bachelor's degree in computer science, electrical engineering, information systems, mathematics, or statistics. Moreover, 53.72% held master's degrees, and 22.87% completed Ph.D. programs.

## 4.2 Model Deployment and Monitoring evaluation

In the survey, we used the same abstraction of seven generic life cycle phases of a popular Brazilian textbook on software engineering for data science [25]: problem understanding and requirements, data collection, data pre-processing, model creation and training, model evaluation, model deployment, and model monitoring. These phases were abstracted based on the nine ML life cycle phases presented by Amershi *et al.* [26] and the CRISP-DM industry-independent process model phases [27]. We asked about the perceived relevance and difficulty of each of the seven phases. For the purposes of this paper and for the sake of simplicity, we represent only the deployment and monitoring life cycle phases.

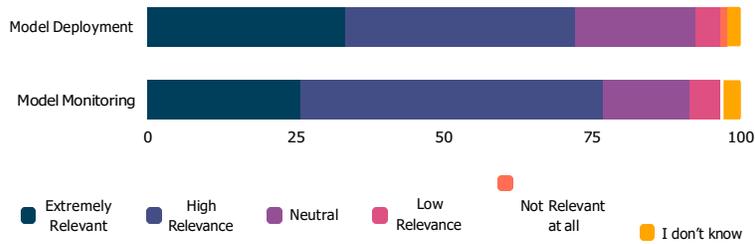

**Fig. 2.** Perceived relevance percentages of the Model Deployment and Model Monitoring activities according to survey participants

The relevance evaluation in Figure 2 shows that the majority of respondents view these activities as highly to extremely relevant, it signifies the critical role they play in the software development life cycle, but still open to an increase in their value for projects.

Although respondents find those relevant, it does not necessarily reflect the expectations with the difficulty represented in Figure 3, where the minority of practitioners find it complex up to very complex, possible due to the new solutions that come with a complete platform ready to have models deployed and, consequently, getting monitored out of the box.

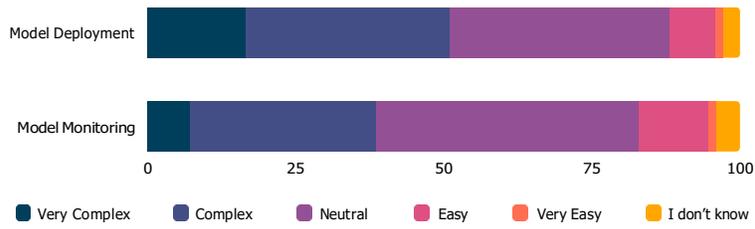

**Fig. 3.** Perceived difficulty percentages of Model Deployment and Model Monitoring activities according to survey participants

### 4.3 What are contemporary practices for deployment? (RQ1) [RQ1.1]

**What kind of approaches are used to deploy ML models?**
For the first question of the survey regarding deployments, the participant were asked about which approach they usually take for hosting their models as shown in Figure 4, where respondents could select more than one option. For the most part, *Service* was the top choice with **P = 59.457 [59.219, 59.695]**, followed by *Embedded Models* with **P = 42.719 [42.476, 42.962]** and *PaaS* with **P = 23.826 [23.628, 24.024]**. Other solutions were also opened for answers and grouped in *Others* with **P = 5.47 [5.359, 5.58]**.

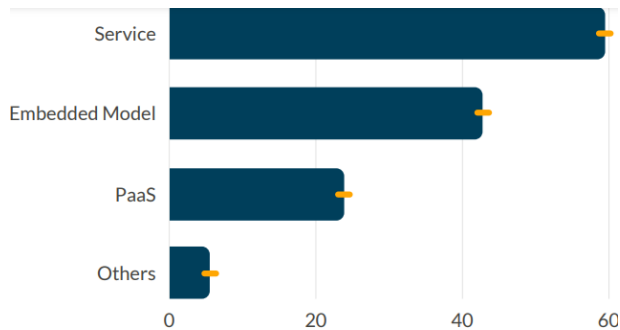

**Fig. 4.** Percentage of deployment approaches used by survey participants (N=168)

**[RQ1.2] Which tools are used for automating model retraining? and [RQ1.3] What are the MLOps practices and principles used?** To de- scribe the usage of MLOps in the life cycle, we asked if the respondents' organi- zations follow any of the practices or principles, followed by a follow up question if a foundational practice, such as an automated retraining pipeline, was used. The results are summarizsed in Figure 5. The majority answered *No* with **P = 70.911 [70.694, 71.128]** and, followed by *Yes* with **P = 29.089 [28.872, 29.306]**. In regards to MLOps, some of the answers were between having their own pipeline built on top a continuous delivery tool (e.g. Gitlab CI/CD [28] and

Azure DevOps [29]) and Machine Learning specific development platform such as BentoML [30], MLFlow [31] and AWS Sagemaker MLOps [32], which follows practices as model re-training and monitoring of relevant aspects.

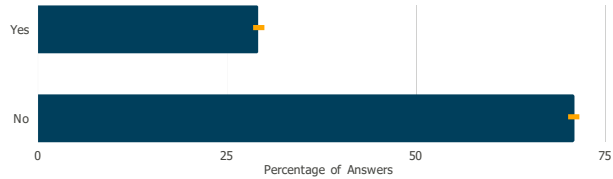

**Fig. 5.** Answers regarding the survey participant's organization usage of MLOps principles (N=168)

## 4.4  What are the main problems faced during the deployment in the ML life cycle stage? (RQ2)

The survey had two questions regarding the main problems faced by practitioners through the deployment and monitoring of models. Figure 6 presents the results of the open and axial coding of the answers for the deployment phase using the probabilistic cause-effect diagrams introduced by Kalinowski *et al.* [33, 34].

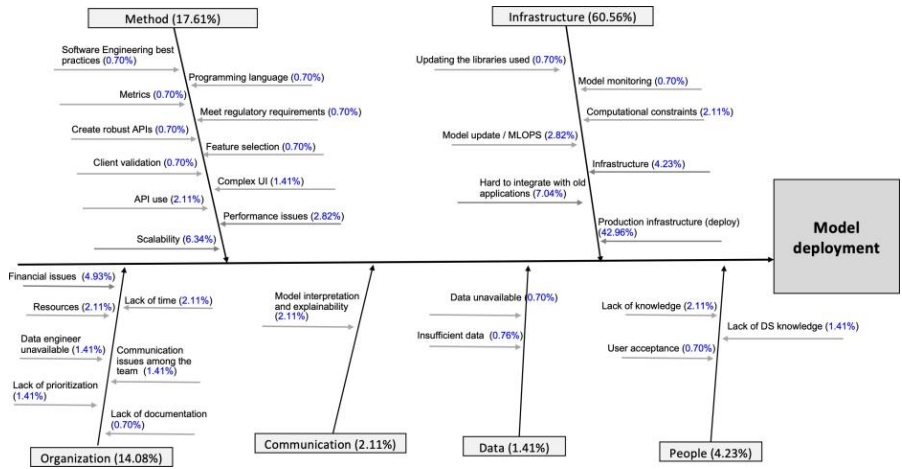

**Fig. 6.** Probabilistic cause-effect diagram related to answers regarding the main problems faced during the model deployment stage (N=142)

As per the survey respondents, the top problems faced within the deployment phase were preparing the infrastructure for production deployment, the difficulty on integrating with legacy applications, what infrastructure architecture to use, how to scale it, and the financial limitations.

### 4.5  What are contemporary practices for monitoring? (RQ3)

**[RQ3.1]  What percentage of the ML-enabled system projects that get deployed into production have their ML models actually being moni- tored?** To evaluate if the deployed projects went through the whole life cycle up until getting monitored, Figure 7 shows that **P = 33.079 [32.842, 33.316]** participants responded that less than 20% of projects do get into production with their aspects monitored, followed by **P = 21.143 [20.942, 21.344]** responding from 20% to 40%, **P = 19.13 [18.943, 19.317]** answering that 80% to 100%, **P = 18.64 [18.456, 18.824]** from 40% to 60% and, finally, **P = 8.009 [7.874, 8.144]** with 60% to 80% get the released project monitored somehow.

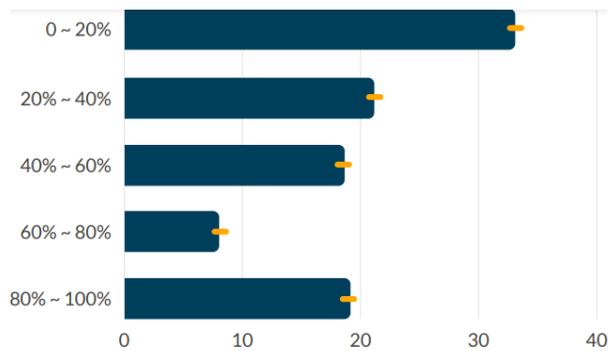

**Fig. 7.** Percentage of answers for models, deployed to production, that have their aspects monitored (N=160)

**[RQ3.2]  What aspects of the models are monitored?** Concerning the model monitoring, respondents described which monitoring aspects were actu- ally monitored as in Figure 8. Participants could be selecting more than one op- tion, having *Input and Output* as the most frequent response with **P = 62.675 [62.431, 62.918]**, followed by *Output and Decisions* with **P = 62.082 [61.834, 62.331]**, *Interpretability Output* with **P = 28.034 [27.805, 28.263]**, *Fairness* with **P = 12.965 [12.792, 13.138]**, and other aspects that were grouped in *Others* with **P = 5.874 [5.761, 5.987]**.

### 4.6  What are the main problems faced during the monitoring in the ML life cycle stage? (RQ4)

Figure 9 presents the results of the open and axial coding of the answers for the main problems of the monitoring phase.

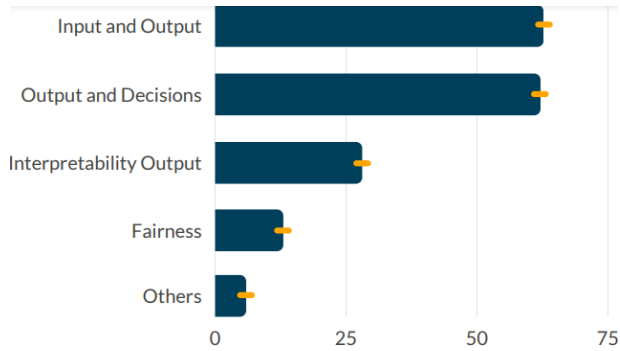

**Fig. 8.** Percentage of answers regarding which of the ML system aspects are monitored (N=153)

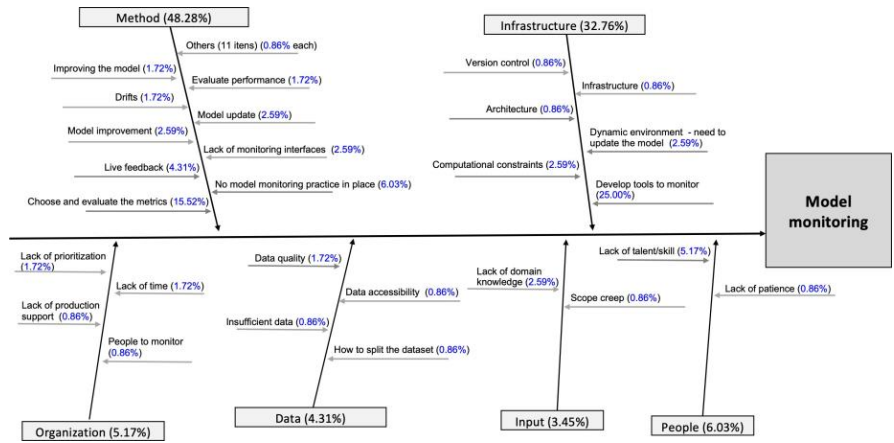

**Fig. 9.** Probabilistic cause-effect diagram related to answers regarding the main problems faced during the model monitoring stage (N=116)

Here, the most observed concerns were related to the need of developing their own monitoring tools, evaluating and choosing the appropriate metrics, while not having any experience in monitoring models and in building monitoringplatforms.

## 4.7  What is the percentage of projects that do go into production? (RQ5)

To describe the population of projects that live up until their general release, data from the demographic questions D7 and D8 (after data cleaning) were combined into Figure 10. As this figure shows, **P = 24.965 [24.759, 25.171]** participants responded that between only 0% to 20% projects went into production, followed by **P = 23.553 [23.337, 23.768]** saying 40% to 60%, then **P =**

**21.221 [21.029, 21.412]** with 80% to 100%, **P = 17.796 [17.618, 17.974]** saying 20% to 40% and, finally **P = 12.465 [12.306, 12.624]** responding with 60% to 80%. By getting all of the percentages calculated and returning the mean value, this leaves us with an average of 45.41% of executed projects reaching gen-eral availability.

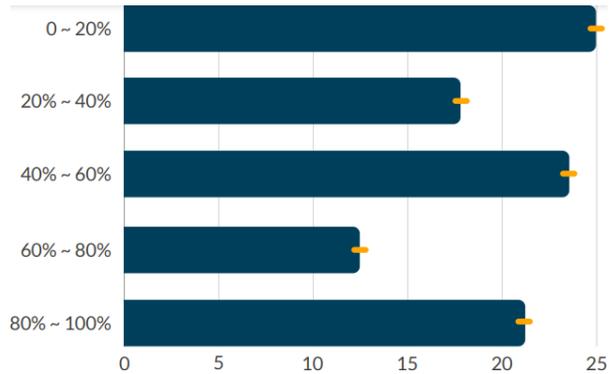

**Fig. 10.** The percentage of ML projects that do go into production (N=169)

## 5 Discussion

Deploying Machine Learning models into production environments can be a complex and challenging task, often accompanied by several problems and considerations. As observed by the survey results as well, the model deployment and monitoring phases are found to be relevant by almost 75% of respondents, corroborating the importance of releasing it to the public and the constant performance analysis for a continuous increase on quality. Although to be found important, its difficulty rates decreased to almost 50% for deployment and 30% for monitoring, showing that a lack of opportunity to evaluate a model that is deployed into production could influence the entire development process analysis. For this case, Mäkinen *et. al.* [35] surveyed data scientists to observe which type of organization would benefit from the MLOps practices, categorizing some of them as the top beneficiaries where the need for model retraining and deployment were extremely important to their natural next step into production models, showing a potential shift in the evaluation if more automated processes were applied to projects.

Through the deployment practices identified, it is evident that ML engineers are deploying most of their models through the Service approach, identifying a growing reliance on cloud-based services that offer comprehensive and scalable solutions already prepared, but compromising customization. Moreover, if integrating with external systems were found to be hard, Embedded Model seemed an alternative approach of choice, leveraging the operation efficiency of an existing software and faster response times, even though its monitoring and scaling difficulty were increased due to the lack of separation from the software that includes the model. At last, having the model deployed in a Platform as a Service approach promises to provide full customization of the infrastructure and flexible environment, although the increasing need for specialized expertise to enable its full potential seems important in this approach through such a complex system.

As per the identified lack of MLOps practices used, participants answered that less than 30% apply some of its principles. This suggests that despite the growing importance of Machine Learning in various industries, a significant number of professionals may not be fully engaged with MLOps, although numerous studies have proven its benefits [36] and providing guidance on establishing the platform [37], unveiling a potential research on how MLOps could influence the work of professionals. Although not fully applied, some of the practices do come embedded in ready to use platforms, also mentioned in the survey, facilitating the adoption quicker than by creating it from the ground up and expanding the usage in seamless way.

To enforce the main problems encountered as per Figure 6, exemplified by this study as issues such as production level infrastructure management and integration with legacy systems, Nahar *et al.* [1] had a systematic literaturereview of challenges in building ML components. They revealed similar results related to deployment, the main challenges encountered along shifts from model-centric to pipeline-driven developments, difficulties in scaling model training and deployment on different types of hardware, and limited technical support for engineering infrastructure. For model monitoring, as per Figure 9, it shows that choosing the metrics and developing new tools to adequate to project's monitoring necessities are the more prominent problems, where Nahar *et al.* observes that the monitorability of a model being considered late to be implemented, providing data quality due to not having well supported tools, lack of support to setup an infrastructure for detecting training-serving skew, and difficulty on designing specific metrics are aligned with the participants' feelings within the survey.

For the monitoring aspects, the survey highlights that the number of models that do go into production and have their aspects monitored are less than 50%, which highlights to us the potential on monitorability exploration for identifying aspects, detecting metrics, and creating new tools to increase the quality attributes of ML models. Following the current status of the monitoring phase, when participants have been asked which aspects were monitored, input and output data stands out. This emphasizes the critical role of data integrity and quality in the overall performance and robustness of Machine Learning systems by identifying potential biases, anomalies, and inconsistencies that could impact the accuracy and reliability of model predictions. Furthermore, monitoring the decisions assesses the correctness and effectiveness of model predictions and the process of decision making to validate the alignment between what was predicted and real-world outcomes. It also shows that the monitoring of interpretability output emerges as another prominent aspect, highlighting the increasing focus on enhancing the transparency and explainability of Machine Learning models, particularly crucial in domains such as establishing trust and verifying model behavior. Lastly, fairness monitoring demonstrates the growing recognition of the ethical implications of algorithms, spurring efforts to monitor and mitigate biases and discriminatory outcomes in model predictions, which underscores the commitment to developing inclusive and equitable Machine Learning systems.

As per Figure 10, less than 50% of projects go into production, still showing a standing pattern where earlier reports [38, 39] and books [40] identified that most of the ML projects fail to get generally available due to several problems. Some of those were identified in this study and are possibly related, such as the organization being unable to fit the infrastructure to the needs of engineering teams, financial issues and not having sufficient expertise on the software engineering process that are, most likely, the lack of specialized professionals. As per Figure 1, qualified personnel such as Cloud Infrastructure Engineers, Data Engineers, and Software Architects were not significantly identified in the team. However, due to the increasing value given to ML models deployment into production, articles such as Heymann *et. al.* [41] will be in evidence to set a common place for frameworks, guides, and books responsible for developing production level ML models and how to apply them.

## 6 Threats to Validity

We identified some threats while planning, conducting, and analyzing the survey results. Hereafter, we list the most prominent threats organized by the survey validity types presented in [42].

**Face and Content Validity**. Face and content validity threats include bad instrumentation and inadequate explanation of the constructs. To mitigate these threats, we involved several researchers in reviewing and evaluating the questionnaire with respect to the format and formulation of the questions, piloting it with 18 Ph.D. students for face validity and with five experienced data scientists for content validity.

**Criterion Validity**. Threats to criterion validity include not surveying the target population. We clarified the target population in the consent form (beforestarting the survey). We also considered only complete answers (*i.e.*, answers of participants that answered all survey sections) and excluded participants that informed having no experience with ML-enabled system projects.

**Construct Validity**. We ground our survey's questions and answer options on theoretical background from previous studies [43, 21] and readings based on identified challenges in model deployment and monitoring [3] and in software architecture [5]. A threat to construct validity is inadequate measurement procedures and unreliable results. To mitigate this threat we follow recommended data collection and analysis procedures [19].

**Reliability**. One aspect of reliability is statistical generalizability. We could not construct a random sample systematically covering different types of pro-fessionals involved in developing ML-enabled systems, and there is yet no generalized knowledge about what such a population looks like. Furthermore, as a consequence of convenience sampling, the majority of answers came from Europe and South America, most of it from Brazil. Nevertheless, the experience and background profiles of the subjects are comparable to the profiles of ML teams as shown in Microsoft's study [44], showing that the nationality attribute did not interfered with the results. To deal with the random sampling limitation, we used bootstrapping and only employed confidence intervals, conservatively avoiding null hypothesis testing. Another reliability aspect concerns inter-observer reliability, which we improved by including independent peer review in all our qualitative analysis procedures and making all the data and analyses openly available online.

## 7 Conclusion

The current study sought to provide a comprehensive overview of the prevailing trends on practices and challenges in model deployment and monitoring within the context of Machine Learning. Through our questionnaire-based online survey targeting practitioners, we identified several key insights allowing us to elaborate as well on potential directions for future research and development. Our analysis underscores the increasing approach on leveraging cloud-based services for model deployment, with a notable emphasis on scalability, accessibility, and seamless integration. This should support the growing demand for efficient and user-friendly deployment solutions, catering to the diverse needs and constraints of contemporary applications. Furthermore, the emphasis on monitoring aspects reflects the heightened awareness of the critical role played by data quality, model accuracy, and transparency in ensuring the reliability and ethical soundness of Machine Learning models.

While the current work provides a comprehensive snapshot of the status quo, it also points towards several areas for further investigation and development. The increasing complexity of Machine Learning models and the dynamic nature of real-world applications, necessitate a more nuanced understanding of deployment and monitoring strategies that can adapt to diverse use cases and evolving challenges. Future research endeavors should prioritize the development of robust and scalable deployment frameworks that accommodate a wide range of ML models and their applications, focusing in better specific infrastructure management and seamless integration to other services. Additionally, there is a pressing need to advance methodologies for comprehensive and real-time monitoring, through incisive metrics discovery and ML-ready monitoring tools, enabling stakeholders to proactively identify and address potential biases, vulnerabilities, and performance bottlenecks in Machine Learning models.

The findings presented in this study contribute to the broader discourse surrounding the deployment and monitoring of Machine Learning models, highlighting the significance of holistic and adaptive approaches that prioritize reliability, interpretability, and observability. By leveraging the insights gleaned from this research, stakeholders and practitioners can take their efforts towards the responsible and impactful development of Machine Learning technologies and researchers can better root their ongoing research on practically relevant needs.